\documentclass[10pt,journal,compsoc]{IEEEtran}

\ifCLASSOPTIONcompsoc
  \usepackage[nocompress]{cite}
\else
  \usepackage{cite}
\fi

\usepackage{afterpage}
\usepackage{amssymb}
\usepackage{booktabs}
\usepackage{cite}
\usepackage{enumitem}
\usepackage[pdftex]{graphicx}
\usepackage{xifthen}
\usepackage{lscape}
\usepackage[framemethod=tikz]{mdframed}
\usepackage{pdflscape}
\usepackage{rotating}
\usepackage{stfloats}
\usepackage{tabularx, xtab}
\usepackage{tikz}
\usepackage{tipa}
\usepackage[hidelinks]{hyperref}
\usepackage{bookmark}

\usepackage[]{todonotes}

\renewcommand{\quote}[1]{\emph{"#1"}}

\begin{document}

\setlist{nolistsep}

\title{Uncovering the Benefits and Challenges of Continuous Integration Practices} 
%{The Many Faces of Continuous Integration: Perceptions and Trade-offs}}
\author{Omar Elazhary, Colin Werner, Ze Shi Li, Derek Lowlind, Neil A. Ernst and Margaret-Anne Storey}
\markboth{IEEE Transactions on Software Engineering}%
% {The Many Faces of Continuous Integration: Perceptions and Trade-offs}
% Peggy trying a new title!
{Uncovering the Benefits and Challenges of Continuous Integration Practices}

\IEEEtitleabstractindextext{%
	\begin{abstract}
		In 2006, Fowler and Foemmel defined ten core Continuous Integration (CI) practices that could increase the speed of software development feedback cycles and improve software quality.
Since then, these practices have been widely adopted by industry and subsequent research has shown they improve software quality.
However, there is poor understanding of \textbf{how} organizations implement these practices, of the \textbf{benefits} developers perceive they bring, and of the \textbf{challenges} developers and organizations experience in implementing them.
In this paper, we discuss a multiple-case study of three small- to medium-sized companies using the recommended suite of ten CI practices. 
Using interviews and activity log mining, we learned that these practices are broadly implemented but \emph{how} they are implemented varies depending on their perceived benefits, the context of the project, and the CI tools used by the organization. 
We also discovered that CI practices can create new constraints on the software process that hurt feedback cycle time.
For researchers, we show that how CI is implemented varies, and thus studying CI (for example, using data mining) requires understanding these differences as important context for research studies.
For practitioners, our findings reveal in-depth insights on the possible benefits and challenges from using the ten practices, and how project context matters.

	\end{abstract}

}

\maketitle

\IEEEdisplaynontitleabstractindextext

% \fi
\IEEEpeerreviewmaketitle

\section{Introduction}
\label{sec:intro}

Continuous Integration (CI) is the practice of iteratively committing small changes into a project's codebase that allows multiple distributed developers to contribute to the same code base and keep their working copy of that code as up to date as possible. CI also supports a number of tools and processes to ensure rapid feedback among developers and help improve code quality. 
In 2006, Fowler and Foemmel \cite{fowler2000continuous} defined the following ten core CI practices and claimed that their use would increase the speed of software development feedback cycles and improve software quality:
 \begin{enumerate}
    \item Maintain a single source repository;
    \item Automate the build;
    \item Make your build self-testing;
    \item Everyone commits to the mainline every day;
    \item Every commit should build the mainline on an integration machine;
    \item Keep the build fast;
    \item Test in a clone of the production environment;
    \item Make it easy for anyone to get the latest executable;
    \item Ensure that system state and changes are visible; and
    \item Automate deployment\footnote{The inclusion of automated deployment with the above practices is nowadays referred to as continuous deployment, however, as we are using Fowler and Foemmel's 2006 post to identify the underlying CI practices, we will be using their terminology as well.}.
\end{enumerate}
 
These practices have been widely adopted in industry, and research has shown that at least some of them do improve software quality, specifically automating builds and self-testing \cite{vasilescu2015quality}, frequently committing to the mainline~\cite{staahl2019big}, and automating deployment~\cite{rossi2016continuous}. Research has also revealed that use of CI decreases integration problems~\cite{staahl2019big}, ensures rapid feedback~\cite{rossi2016continuous}, increases software quality~\cite{vasilescu2015quality}, and improves developer productivity \cite{hilton2016usage}.

Most of the existing research on CI has focused on mining repositories to study the impact of CI on software quality. It has not addressed understanding how the ten CI practices each play a role in quality improvement or developer productivity.
There is also poor understanding of \textbf{how} or \textbf{if} organizations implement the individual practices (only four practices have been extensively studied), of the \textbf{benefits} developers perceive from adopting them, of the \textbf{challenges} developers and organizations experience when implementing them, 
 and how \textbf{project context} plays a role. 

But why does this lack of knowledge about CI matter? For teams and organizations, they must decide which practices to use given their context and resources, and the expected benefits and drawbacks each practice brings.
Consider the practice of keeping the build fast.
At the very least, this requires a commitment to highly modular code, reduced code size, and/or expensive build machines.
And it likely requires dedicated build engineers to optimize compilers and linkers.
Developers and managers also need to decide exactly what ``keep the build fast'' implies: Overnight? Within an hour? 
And these decisions are markedly different for a system with a million lines of legacy code than with a small project in a start-up environment.

To help address this lack of knowledge, we 1) investigate the extent to which three software organizations implement the ten CI practices, 2) the benefits they experience as a result of these practices, and 3) the challenges faced in \emph{their implementation}. 
To answer our questions, we conducted a multiple-case study with mixed methods at three small- to medium-sized software-as-a-service organizations.
First, we performed data-driven analysis of the CI artifacts of the three case companies, and then we conducted interviews with members of each company. 
The selection of the three companies was opportunistic as some of the authors had previously conducted studies with them. 
These existing relationships meant we had established trust with the organizations, which was necessary to access internal documents and artifacts. More importantly, our relationships helped us interpret the data we collected using contextual knowledge of the companies and their processes.

From our case studies, we learned which of the ten CI practices the three companies adopted, how and why they adopted them, the benefits they perceived, and the challenges they encountered.
We discovered that the three organizations differ in how they adopt and implement these practices because of \textbf{project context}, \textbf{practice perceptions}, and \textbf{process constraints}.
% Peggy: isn't this next sentence redundant?
% Consequently, these factors influenced the way CI practices were implemented, resulting in different implementations of the same practice across organizations.
Researchers should consider these key differences in how CI is adopted when they study projects that report a use of CI. 
Our findings also inform practitioners that these context factors matter, and what benefits and challenges they may expect.
For example, CI itself can constrain development time, and adopting CI practices requires organization-specific definitions of important CI metrics, such as build time.

The rest of this paper is organized as follows. In Section \ref{sec:background}, we discuss CI concepts and then map the ten CI practices to benefits and challenges reported in the literature.
Then, Section \ref{sec:study} showcases our study design, the methods used, and how we incorporated qualitative data analysis to complement the observations from our data-driven analysis.
Section \ref{sec:findings} reports each organization's implementation of the ten CI practices, and Section \ref{sec:discussion} emphasizes the main findings from these profiles.
Finally, Section \ref{sec:threats} discusses the limitations that apply to our study.

\section{Background: Continuous Integration Practices}
\label{sec:background}

CI is the practice of iteratively committing small changes into a project's codebase in order to keep each developer's working copy of that code as similar as possible \cite{fowler2006continuous, fitzgerald2017continuous}.
To help facilitate automation within CI, special CI tools are typically used (e.g., Jenkins, Travis).
There are many CI tools available and organizations typically use a variety of different tools \cite{guerriero2019adoption}.

Regardless of which tool is used, the tools collectively create an automated \emph{build pipeline} that contains a number of stages depending on the organization's needs (e.g., building, testing, deployment, etc.).
The build pipeline supports and enables an organization to make small, frequent code changes that are regularly integrated, tested, and included in the product \cite{fitzgerald2017continuous, shahin2017continuous}.

However, CI is more than just the use of an automated build system or tool that facilitates automation.
Viewing CI as a set of practices as opposed to tools is corroborated by M{\aa}rtensson et al. \cite{maartensson2017continuous}, who showed that the build system on its own is just one of many factors related to (and often causing) undesirable CI behavior.
Additionally, while the ten CI practices advocate the need for testing (e.g., practice 3 mentions that builds should be self-testing), they do not prescribe a particular level of test quality.
The guideline simply mentions that imperfect tests which are run frequently are better than perfect tests which are never executed.
CI also includes guidelines for development pace and workflow transparency.
However, the existence of these guidelines and proposed practices does not guarantee that their application is consistent, nor that they produce the desired effects.

Researchers and practitioners have long attributed several benefits to CI,
and Fowler and Foemmel \cite{fowler2006continuous} (and later Humble and Farley \cite{humble2010continuous}) do not mention any possible risks from implementing continuous practices.
The systematic literature reviews on CI practices do mention several impediments to adoption \cite{laukkanen2017problems} but do not say much about the risks from using the practices.
Practitioners have indicated there is some pushback against CI, but still recommend CI adoption as an overarching practice, rather than decomposing it into the ten individual practices that comprise CI~\cite{berg2019common, heller2015continuous, bugayenko2014continuous}.

We review the ten CI practices as they are defined in the literature, and the main benefits and challenges reported.
The ten practices, their benefits and challenges are also summarized in Table \ref{tbl:practices_lit}.

\begin{table*}[!t]
    \caption{Literature mapping of practices to claimed benefits and possible challenges}
    \label{tbl:practices_lit}
    \centering
    \begin{tabularx}{2\columnwidth}{m{4.46cm} m{6cm} m{6cm}}
        \toprule[1pt]
        \textbf{Practice} & \textbf{Claimed Benefits} & \textbf{Possible Challenges} \\
        \toprule[1pt]
% P1: Maintain a single source repository
        Maintain a single source repository &
        \begin{itemize}
        	\item Facilitates project building \cite{fowler2006continuous}
        	\item Enhanced codebase visibility \cite{jaspan2018advantages}
        	\item Centralized dependency management \cite{jaspan2018advantages}
        \end{itemize} &
    	\begin{itemize}
    		\item Slower build speed due to dependency building \cite{jaspan2018advantages}
    	\end{itemize} \\
    	\midrule
% P2: Automate the build
        Automate the build &
        \begin{itemize}
        	\item Reduces mistakes resulting from manual steps \cite{fowler2006continuous}
        	\item Higher commit rate \cite{zhao2017impact}
        	\item Higher pull request processing rate \cite{zhao2017impact}
        	\item Higher software quality \cite{vasilescu2015quality}
        \end{itemize} &
    	\begin{itemize}
    		\item Build complexity increases \cite{laukkanen2017problems, duvall2011continuous}
    	\end{itemize} \\
    	\midrule
% P3: Make the build self-testing
        Make the build self-testing &
        \begin{itemize}
        	\item Faster and efficient bug capture \cite{fowler2006continuous, staahl2013experienced, hilton2016usage}
        \end{itemize} &
        \begin{itemize}
        	\item Inadequate testing strategies \cite{shahin2017continuous}
        	\item Ambiguous test results \cite{laukkanen2017problems}
        	\item Flaky tests \cite{laukkanen2017problems}
        	\item Manual testing reduces utility from CI \cite{duvall2011continuous}
        \end{itemize} \\
    	\midrule
% P4: Everyone commits to the mainline every day
        Everyone commits to the mainline every day &
        \begin{itemize}
        	\item Quick conflict detection and resolution \cite{fowler2006continuous, staahl2019big}
        \end{itemize} &
        \begin{itemize}
        	\item Large commits and long-lived branches cause merge conflicts \cite{laukkanen2017problems, duvall2011continuous}
        \end{itemize} \\
    	\midrule
% P5: Every commit should build the mainline on an integration machine
        Every commit should build the mainline on an integration machine &
        \begin{itemize}
        	\item Ensures new changes do not break existing functionality or introduce new bugs \cite{fowler2006continuous}
        \end{itemize} &
        \begin{itemize}
        	\item Work blockage may occur due to merge conflicts \cite{laukkanen2017problems, duvall2011continuous}
        \end{itemize} \\
    	\midrule
% P6: Keep the build fast
        Keep the build fast &
        \begin{itemize}
        	\item Provides rapid feedback \cite{fowler2006continuous}
        	\item Reduces effects of context switching \cite{hukkanen2015adopting}
        \end{itemize} &
        \begin{itemize}
        	\item Work blockage as a result of lengthy and complex builds \cite{laukkanen2017problems}
        \end{itemize} \\
    	\midrule
% P7: Test in a clone of the production environment
        Test in a clone of the production environment &
        \begin{itemize}
        	\item Recognize problems in production early before deployment \cite{fowler2006continuous}
        \end{itemize} &
        \begin{itemize}
        	\item Creating production clone too close to a release date \cite{duvall2011continuous}
        \end{itemize} \\
    	\midrule
% P8: Make it easy for anyone to get the latest executable
        Make it easy for anyone to get the latest executable &
        \begin{itemize}
        	\item Gives users access to feedback before deployment \cite{fowler2006continuous}
        	\item Facilitates agile testing \cite{staahl2013experienced}
        \end{itemize} &
        \begin{itemize}
        	\item Build complexity impacts developer ability to build locally \cite{laukkanen2017problems}
        	\item Developers have to perform manual build steps \cite{duvall2011continuous}
        \end{itemize} \\
    	\midrule
% P9: Ensure that system state and changes are visible:
        Ensure that system state and changes are visible &
        \begin{itemize}
        	\item Facilitates communication between project team members \cite{fowler2006continuous, staahl2013experienced}
        \end{itemize} &
        \begin{itemize}
        	\item Not sending or ignoring build notifications \cite{duvall2011continuous}
        	\item Build information and notifications are not available to the entire team \cite{duvall2011continuous}
        	\item Lack of discipline when monitoring build status \cite{laukkanen2017problems}
        \end{itemize} \\
    	\midrule
% P10: Automate deployment
        Automate deployment &
        \begin{itemize}
        	\item Reduces errors resulting from manual steps \cite{fowler2006continuous}
        	\item Speeds up the deployment process \cite{fowler2006continuous}
        	\item Faster release cycle \cite{hilton2016usage}
        \end{itemize} &
        \begin{itemize}
        	\item Deployment complexity affects its success \cite{laukkanen2017problems, duvall2011continuous}
        \end{itemize} \\
        \bottomrule[1pt]
    \end{tabularx}
\end{table*}
\vspace{2.5mm}

%%%%
\subsection{Maintain a Single Source Repository}
Fowler recommends using a single source repository, with the expected benefit that doing so would reduce friction in identifying and accessing relevant sources and tools \cite{fowler2006continuous}.
%In theory, a developer could check out the repository on a machine with the bare minimum tools required, and be able to build it with minimal effort.
Other claimed benefits include enhanced codebase visibility since all of the application code---and the API---is in a central location, reducing cognitive load for others to understand changes, as Jaspan et al. found \cite{jaspan2018advantages}.
%Developers who report a reduction in cognitive load report that it is due to the presence of the API within the same repository as the source code, which facilitates understanding the latter.
However, research that has studied CI has focused primarily on open source projects \cite{vasilescu2015quality} where a project's structure is generally oriented towards a microservice architecture with dependencies that are typically separated from an application's source code \cite{kikas2017structure}.
Studies that involve projects in industry 
% also do not clarify 
%or discuss repository structure; for instance, they 
do not indicate what repository structure is being used or whether it has an impact on the findings \cite{rossi2016continuous}.

Some challenges from this practice are mentioned in the literature. Jaspan et al. \cite{jaspan2018advantages} mention that a single source code repository may result in a lengthy build time. 
Zampetti et al. \cite{zampetti2020empirical} also bring up issues related to how project repositories are structured, but these issues are related to how project code is structured from a software design perspective as opposed to build complexity.

%%%%
\subsection{Automate the Build}
Fowler and Foemmel \cite{fowler2006continuous} claim that automating the build will reduce mistakes from manual steps.
Another benefit observed by Zhao et al. \cite{zhao2017impact} is the increase in commit frequency as displayed by commit and pull request rates, which contributes to increased development velocity.
Hilton et al. \cite{hilton2016usage} also note that their participants claimed the use of build automation produces higher quality software.

Laukkannen et al. \cite{laukkanen2017problems} and Duvall \cite{duvall2011continuous} report that a complex build may be a barrier to using CI. 
In Laukkannen's case, a build that contained several special cases to accommodate multiple teams working on the product made the build so complex that it became harder to change later on.

%%%%%
\subsection{Make your Build Self-Testing}
The claimed benefit of self-testing builds is that it should catch bugs quickly and efficiently. This was corroborated by two studies \cite{staahl2013experienced,hilton2016usage}.
St{\aa}hl and Bosch \cite{staahl2013experienced} report that their respondents found the use of automated builds improves project predictability as it helps to identify problems faster.
Hilton et al. \cite{hilton2016usage} report that their respondents found the tests used within an automated build help catch bugs.
However, Fowler and Foemmel \cite{fowler2006continuous} note that this efficient bug-catching is a function of the tests (or test suite) employed, as opposed to something inherent to continuous practices.

Most of the challenges reported with this practice arise from the lack of proper testing strategies and poor test quality as reported by Shahin et al. \cite{shahin2017continuous}.
%Tests need to be planned out, given adequate infrastructure resources, and effort needs to be invested into test labor.
%These may be impediments to implementing this practice.
Test quality, flaky tests \cite{laukkanen2017problems}, low test coverage, and long-running tests are also risks from partially implementing this practice.  

%%%%%
\subsection{Everyone Commits to the Mainline Every Day}
This practice claims to facilitate quick conflict detection and resolution because when every developer commits to the mainline (master branch) often, the divergence becomes small enough to easily resolve integration conflicts \cite{fowler2006continuous} and offers a way to guarantee that tests, no matter how imperfect they may be, are run frequently.
Additionally, committing smaller chunks helps track productivity and provides developers with a sense of progress \cite{fowler2006continuous}.

However, St{\aa}hl et al. \cite{staahl2019big} found that code properties (i.e., complexity) impact whether developers commit more often---the more complex the area of code being changed, the more likely developers are to commit often.
Laukkannen et al. \cite{laukkanen2017problems} report that merge conflicts are a common problem because of long-lived branches (where developers have worked in isolation for an extended period of time) or large commits.
Laukkannen et al. \cite{laukkanen2017problems} as well as Shahin et al. \cite{shahin2017continuous} report on merge conflicts causing work blockage.
%While work blockages can be due to a slow approval process for commits, they may also be caused by a complicated dependency structure.

%%%%% 
\subsection{Every Commit Should Build the Mainline on an Integration Machine}
Fowler and Foemmel \cite{fowler2006continuous} claim that triggering builds at the commit level is comparable to frequently running tests, which helps ensure that new changes do not introduce bugs or break existing functionality.

There is some research that indicates commit size and change complexity are factors for predicting build success, and that
larger, more complex changes may have a negative impact on build success \cite{islam2017insights}.
Similarly to practice 3 (make the build self-testing), the level at which the build runs is highly dependent on resources~\cite{shahin2017continuous}.
Determining the level at which a build should trigger and how many resources should be allocated to it may be impediments to adoption of this practice.

\subsection{Keep the Build Fast}
This practice is claimed to provide developers with rapid feedback \cite{fowler2006continuous},
which has been shown as a benefit \cite{rossi2016continuous} that fits the current trend of agile, high-velocity software development as indicated by Fitzgerald and Stol \cite{fitzgerald2017continuous}.
Additionally, Hukkannen \cite{hukkanen2015adopting} mentions that keeping the build fast decreases context switching between tasks which may happen when developers have to wait for slow build results.  
Fast builds can also mitigate potential work blockages~\cite{laukkanen2017problems}.

%%%%% 
\subsection{Test in a Clone of the Production Environment}
Testing in an environment that is as close to the production environment as possible allows developers to recognize problems before deployment~\cite{fowler2006continuous}.
To our knowledge, there has been little research on this practice. 
While no challenges or adoption impediments have been reported, it is conceivable that issues related to testing strategy and flaky tests across diverse environments can also deter its adoption. 

%%%%%
\subsection{Make It Easy for Anyone to Get the Latest Executable}
Early access to the final product (e.g., executable, binary, etc.) allows users to give feedback before deployment \cite{fowler2006continuous}.
It also allows developers to try out new features to do exploratory testing and stay up to date with the latest application status.
St{\aa}hl and Bosch \cite{staahl2013experienced} report that easy access to an executable 
facilitates agile testing by a non-technical client. 

If a developer has to perform several manual commands to prepare their environment for local builds, this may signal that this practice is only partially applied \cite{duvall2011continuous}.
Build complexity, as reported by Laukkannen et al. \cite{laukkanen2017problems}, can
keep a developer from building locally. 

%%%%%
\subsection{Ensure that System State and Changes Are Visible}
St{\aa}hl and Bosch \cite{staahl2013experienced} confirm that through the use of build system interfaces or dashboards, visible system state and changes ease communication between both co-located and remote project team members~\cite{fowler2006continuous}.
However, St{\aa}hl and Bosh do not elaborate on how developers ensure the system and its changes are visible.
Duvall \cite{duvall2011continuous} reports that unsent notifications, ignored notifications, or otherwise uninformative notifications are likely signals that this practice is not being fully implemented.

%%%%%
\subsection{Automate Deployment}
Similarly to automating the build, this practice reduces errors introduced by manual steps and speeds up the deployment process~\cite{fowler2006continuous}. 
Automated deployment is nowadays referred to as continuous deployment, however, as we are using Fowler and Foemmel's 2006 post to identify underlying CI practices, we will be using their terminology.
Automated deployment leads to faster release cycles in projects that use an automated build tool against those that do not~\cite{hilton2016usage}.
Hilton et al. found that the projects that automate their deployments release about twice as often as those that do not.
This increased deployment rate allows for more frequent customer feedback \cite{rossi2016continuous}.

Laukkannen et al.~\cite{laukkanen2017problems} identify deployment build complexity as a possible risk: the more complex the deployment, the harder it is to ensure its success.
This risk grows in importance if customers cannot tolerate downtime.

\subsection{Summary}
Existing research has explored the expected and experienced benefits from using CI in general, as well as some adoption challenges and perceived risks from using the specific ten practices. 
However, we do not know if all ten practices are needed to obtain the expected benefits, nor do we know what trade-offs may need to be considered in their adoption.
%This kind of nuanced information is hard to obtain from examining trace data alone and thus the focus of our study. 

\section{Methodology}
\label{sec:study}

We conducted a multiple-case study with three software development organizations and posed the following three research questions:  
\vspace{-1mm}
\begin{mdframed}
   \begin{enumerate}[label=\textbf{RQ\arabic*:},leftmargin=*]
       \item \textbf{Which CI practices} do these organizations apply, and how do they apply them? 
       \item What \textbf{benefits} do these organizations attribute to the use of these CI practices?
       \item What \textbf{challenges} do they experience adopting or using these CI practices?  
   \end{enumerate}
\end{mdframed}
\vspace{-1mm}

%\subsection{Studied Organizations}

% Why did we go mixed methods
% > Quantitative is how literature has made conclusions about continuous practices
% > Qualitative adds context around the quantitative data we observe (the WHY)
We recruited three organizations who self-identified as continuous, our main inclusion criterion for the case studies~\cite{yin2003case}.
For each case, we used a mixed-methods strategy \cite{storey2020software}, combining interviews and trace-log analysis.
Our unit of analysis was an organization where we studied how they use the ten continuous integration practices defined by Fowler and Foemmel~\cite{fowler2006continuous}.
The selection of the organizations was opportunistic as we approached local companies we could visit in person (to gain context through prolonged visits) and who trusted us to analyze their data, conduct interviews, and with whom we could confirm our findings. 
These organizations also perceived our study to be of value to them as we would help them reflect on how and why they use CI practices.

At the time of our study, each organization was roughly 8 years old and had 30-60 employees. 
We refer to the organizations as A, B, and C in the remainder of the paper.
Organization A is involved in the data business, and frequently collects and processes large amounts of data.
Organization B is an online content provider and does advertisement management.
Organization C provides an online booking platform that integrates with multiple existing services for many customers worldwide.
All three organizations follow a software-as-a-service (SaaS) model for their products, one of the most common form of cloud applications~\cite{costello2019gartner}.
They also use automated build tools that support continuous practices, and self-identify as continuous (A,B) or becoming continuous (C).
%The three organizations all describe themselves as being continuous and follow agile and continuous practices according to Humble and Farley's book~\cite{humble2010continuous}.
More information on the three organizations can be found in Table \ref{tbl:case_summary}.

\begin{table*}
	\caption{Organization Characteristics}
	\label{tbl:case_summary}
	\centering
	\begin{tabular}{l l l l}
		\toprule[1pt]
		 & Organization A & Organization B & Organization C \\
		\toprule[1pt]
		Domain & Data Processing & Online Content Provider & Online Bookings \\
%		Age & 9 (Founded 2011) & 9 (Founded 2011) & 10 (Founded 2010) \\
		Team(s) Interviewed & 4 & 1 & 3 \\
		Average Team Size & 3-4 & 8-10 & 6-12 \\
		Participants Interviewed & 5 & 5 & 8 \\
		Employee Distribution & Co-located & Distributed & Co-located \\
		Reported Continuous Status & Continuous & Continuous & Moving towards Continuous \\
		\bottomrule[1pt]
	\end{tabular}
\end{table*}

We collected data from several sources for each organization, including developers, managers, development activity logs, and deployment logs (discussed in Section~\ref{sec:activitylogs}).
Within each organization, we first examined activity log data to objectively identify if they used the different CI practices (answering RQ1). 
We then conducted interviews with several developers and one or more managers from each organization.
Our interview participants kindly recruited other participants for us.
The interviews helped us form \textbf{case report profiles} that give a deeper introduction to the three organizations and how they use CI practices.  
The interviews further helped us answer our three research questions. 

\subsection{Development Activity Log Mining}
\label{sec:activitylogs}

We extracted development activity logs and task management data dumps for all three organizations, as well as deployment logs for organization B.
For organization A, we collected data from January 2019 to November 2019, for organization B we collected data from January 2017 to June 2019, and for organization C we collected data from October 2017 to October 2019.
Our goal from this step in our research was to verify that the companies used automated builds and to identify their commit frequency, the speed of builds, and whether commits triggered builds.

We found that only four out of the ten practices could be verified with activity log data alone:
\begin{itemize}
	\item \textbf{Automate the build:} The existence of build logs confirmed that automated builds are used in the three organizations.
	\item \textbf{Everyone commits to the mainline every day:} Although the commit activity was available through the trace logs, we found that some of the commit history was lost as some commits were squashed and others were cherry-picked from a developer's version of the repository. 
	\item \textbf{Every commit should build the mainline on an integration machine:} From the repository activity, we were able to determine if each commit resulted in a build since each commit features a build status indicator.
	\item \textbf{Keep the build fast:} The build logs included timestamps that helped calculate build durations.
\end{itemize}
Although we could determine the use of these four practices, this data does not inform about \emph{why} certain practices are used, if other practices not visible in the trace data are used, whether the used practices produced their claimed benefits, or if the organizations had challenges with these or the non-adopted practices. 
For this reason we conducted on-site interviews.

\subsection{Interviews}
% Why is SaaS important?
We interviewed a cross-section of organizational members (developers, managers, team leads, and directors) from the three organizations. 
Within each organization, we followed an opportunistic recruitment strategy whereby our contact would announce that we were conducting a study on continuous practices and were looking for participants across roles.
This cross-section of 18 subjects helped us determine whether the implementation of continuous development practices had effects beyond the scope of the development process and helped us triangulate our findings across participant accounts (notably developers and managers) and the development logs.
Our participant pool consisted of the following:
\begin{itemize}
	\item 1 senior developer;
	\item 1 data science team lead;
	\item 1 front-end web developer;
	\item 1 project lead;
	\item 3 DevOps engineers;
	\item 2 full-stack developers;
	\item 2 product owners;
	\item 1 chief marketing officer;
	\item 1 director of technical support;
	\item 1 chief technology officer;
	\item 1 account manager;
	\item 1 chief operating officer;
	\item 1 junior developer; and
	\item 1 director of sales.
\end{itemize}
To protect anonymity, we labeled each developer and team lead (highly technical roles) as developers, and the remaining roles as managers, as shown in Table \ref{tbl:interviewees}.

\begin{table}[!t]
    \caption{Participant Mapping to Organization and Role}
    \label{tbl:interviewees}
    \centering
    \begin{tabular}{c c c}
        \toprule[1pt]
        Participant & Role & Organization \\
        \toprule[1pt]
        P1 & Developer & A \\
        P2 & Developer & A \\
        P3 & Developer & A \\
        P4 & Developer & A \\
        P5 & Manager & A \\
        P6 & Developer & B \\
        P7 & Developer & B \\
        P8 & Developer & B \\
        P9 & Manager & B \\
        P10 & Manager & B \\
        P11 & Manager & C \\
        P12 & Developer & C \\
        P13 & Manager & C \\
        P14 & Manager & C \\
        P15 & Manager & C \\
        P16 & Manager & C \\
        P17 & Developer & C \\
        P18 & Manager & C \\
        \bottomrule[1pt]
    \end{tabular}
\end{table}

With these participants, we conducted semi-structured interviews based on the ten core practices within our definition of continuous integration.
For each participant, we first asked them what their definition of continuous practices was, and what impact they perceived those had on their work.
Then, for each of the ten practices in Table \ref{tbl:practices_lit}, we asked each participant whether their organization implemented it and why.
The questions were open-ended in nature, and we asked participants to elaborate on issues relevant to the perceived impact from practice implementation, and why the practice was either adopted or rejected.

At least two interviewers were present to mitigate any bias a single interviewer might introduce.
The interviews were recorded for later transcription and coding.
We also made sure to clarify our understanding of participant responses, even going so far as to rephrase their answers back to them to guarantee we shared the same understanding.
Examples of the questions we asked include:
\begin{itemize}
	\item Are you familiar with continuous practices? Please explain.
	\item For the project with which you interact the most, does all of its source code reside in a single repository? Why/Why not?
	\item For the project with which you interact the most, is there an automated build/test process attached to it? Why/Why not?
	\item For the project with which you interact the most, does every commit to the master branch kick off an automated build/test process? Why/Why not? If not, how often are automated build/test processes triggered on the master branch?
\end{itemize}
A complete list of our starting questions as well as some of the questions that we asked based on participant responses is included in our reproduction package at \url{http://doi.org/10.5281/zenodo.3950115}.

\subsection{Thematic Analysis of Interview Data}
% Coding as described in Dana's paper
% Themes based on the codes
Following a similar approach to Heikkil{\"a} et al. \cite{heikkila2017managing}, we followed an inductive coding strategy, whereby codes emerged organically from the transcripts.
To generate a common codebook, we selected four transcripts to code initially.
Each transcript was coded by a pair of coders following the constant comparison method.
For each passage within a transcript, we assigned a code (a label that would describe its general idea).
If none of the existing codes described the passage adequately, a new code was created to better capture the concept the passage was conveying.
After the first four transcripts were coded, the four coders met to discuss and agree on what codes to use.
Duplicate and overlapping codes were merged to reduce clutter.
The end result of this process was our initial codebook, available in our reproduction package.

For the remaining 14 transcripts, we followed the same method where two coders would code a single transcript and then meet  to compare the results and settle on an interpretation.
Any new codes were added to the codebook, and the codebook was later examined to merge any duplicates.
%Our median inter-rater agreements across coding rounds, as represented by Cohen's kappa, are given in Table \ref{tbl:agreement}.
%
%\begin{table}
%	\caption{Median inter-rater agreement across all four coders}
%	\label{tbl:agreement}
%	\centering
%	\begin{tabular}{r r}
%		\toprule[1pt]
%		 & Coder 1 \\
%		\toprule[1pt]
%		Coder 2 & 0.49 \\
%		Coder 3 & 0.48 \\
%		Coder 4 & 0.49 \\
%		\bottomrule[1pt]
%	\end{tabular}
%\end{table}

Once we coded the transcripts for all 18 participants, we conducted thematic analysis sessions.
We revisited the transcripts and our notes, and listed common themes as supported by the co-occurrence of codes.
Throughout our thematic analysis sessions, we debated the themes and our interpretation of them.
After merging duplicates and overlapping themes, we constructed our final list of themes, which we verified by conducting a member checking survey with the participants.

\subsection{Member Checking}
To verify our interpretation of the themes we found in the interview analysis, we sent a brief survey to our interview participants (this survey is included in our reproduction package). 
For the most common themes that emerged in our thematic analysis, we asked each participant to indicate their agreement with the theme only if it applied to the project with which they were most familiar, using a 5-point Likert scale to record agreement.
Participants were also given the option to select "Not Sure" if they were unsure about whether a particular statement applied to their project.
Eight interview participants completed the entire survey (44\%).

A couple of the themes we identified in the interviews did not see broad agreement from the participants that answered our confirming survey, and thus we removed them from the themes we report in Section \ref{sec:discussion}. 

Using the findings from the data analysis and interviews, and the confirmation of themes we received from the survey, we developed a case report profile for each organization that provides a summary of how each organization uses the CI practices (or not), the benefits reported and any challenges we identified (Section \ref{sec:findings}).
Detailed versions of the case reports are included in the paper's reproduction package.
We asked our main contacts at each organization to forward the organization case report and a copy of our paper to the interviewed participants.
We received positive feedback from one member in each of the three organizations (team lead from A, program manager from B, DevOps engineer from C) about the accuracy of the reports and paper.
They reported only minor discrepancies, which we updated in the reports and the paper.
For instance, our contact at organization A commented that \quote{it seems mostly fine} and clarified that their build infrastructure was hosted on AWS, which was not mentioned during the interviews.
Our contact at organization B was appreciative of our analysis and commented: \quote{Thanks for the recommendations. We continue to make forward progress on most of the items you mention.}
Similarly, our contact at organization C mentioned that our analysis \quote{looks good}.

\vspace{2.5mm}

The limitations of our study can be found in Section \ref{sec:threats}. We present our findings next, which includes a summary of the case report profiles and answers to our three research questions. 
\section{Findings}
\label{sec:findings}
We first describe the case profiles for the organizations that participated in our multi-case study.
We have anonymized each organization and their respective employees to protect their identities.
For each organization, we describe the context in which their projects occur and their characteristics.
Then, we discuss the main \emph{differences} in how the CI practices are used and each company's rationale for using (or not) that practice.
We focus on the four practices with the bigger differences.
A full overview of how practices were used, their benefits, and challenges can be found in Table \ref{tbl:practices_orgs}.

\newcommand{\rowprcheader}[2]{
	\multicolumn{3}{l}{\textbf{#1}}
	\hfill
	\ifthenelse{\isempty{#2}}{\textit{No difference}}{\textit{Differences due to #2}}
}
\newcommand{\roworgheader}[1]{\textit{#1}}
\renewcommand{\arraystretch}{0.8}
\aboverulesep = 0mm
%\belowrulesep = 0.5mm

\begin{table*}[!t]
    \centering
    \caption{Comparing CI practices at organizations A, B, C. We list, per CI practice, how that practice is implemented and rationalized, what trade-offs are perceived, and why its implementation differs (italics). Organization codes in parentheses indicate data from the organization supports the finding.}
    \label{tbl:practices_orgs}
	\begin{tabularx}{2\columnwidth}{p{5.5cm} p{5.5cm} p{5.5cm}}
		\toprule[1pt]
		\textbf{How a Practice is Implemented (RQ1)} & \textbf{Rationale/Benefit (RQ2)} & \textbf{Challenges/Trade-Offs (RQ3)} \\
		\midrule[1pt]
	% P1: Maintain a single source repository
		\rowprcheader{Maintain a single source repository}{practice perception} \\
		\begin{itemize}[leftmargin=*]
			\item Logical project separation across repositories (A).
			\item Microservice-like architecture (B).
			\item Inconsistent grouping of dependencies and main project (C).
		\end{itemize} &
		\begin{itemize}[leftmargin=*]
			\item Minimizes merge conflicts (A).
			\item Mono-Repo rationale not obvious at the time (B).
			\item Mono-Repo tool support not available at the time (C).
		\end{itemize} &
		\begin{itemize}[leftmargin=*]
			\item Boundaries not always clear (A).
			\item Dependency management is complex (B).
			\item Workflow duplication (B).
		\end{itemize} \\
		   	\midrule
		% P2: Automate the build
		\rowprcheader{Automate the build}{project context} \\
		\begin{itemize}[leftmargin=*]
			\item Via Jenkins (A, C).
			\item Via AWS CodePipeline and AWS CodeBuild (B).
		\end{itemize} &
		\begin{itemize}[leftmargin=*]
			\item Consistency (A, B)
			\item Reproducibility (A, B)
			\item Swift development (A).
			\item Delegates repeatable tasks to build (A, C).
		\end{itemize} &
		\begin{itemize}[leftmargin=*]
			\item Build can be a bottleneck (B).
		\end{itemize} \\
		   	\midrule
		% P3: Make the build self-testing
		\rowprcheader{Make the build self-testing}{tool constraints, practice perception, and project context} \\
		\begin{itemize}[leftmargin=*]
			\item Unit tests in PR builds (A, B, C).
			\item Regression tests in mainline builds (A, B, C).
			\item Integration tests in PR builds. Local, PR, mainline, QA, then product manager tests (C).
		\end{itemize} &
		\begin{itemize}[leftmargin=*]
			\item Reliable bug detection (A).
			\item Ensures consistency (B).
			\item Minimizes breaking changes (C).
		\end{itemize} &
		\begin{itemize}[leftmargin=*]
			\item Fast tests due to limited coverage (B).
			\item Limited by infrastructure (B).
			\item UI difficult to test. Cost-Benefit of adding automated UI tests not clear (A).
			\item Longer builds (C).
			\item Decreased velocity (C).
		\end{itemize} \\
		   	\midrule
		% P4: Everyone commits to the mainline every day
		\rowprcheader{Everyone commits to the mainline every day}{practice perception} \\
		\begin{itemize}[leftmargin=*]
			\item Frequent commits on most projects (A).
			\item Daily commits are norm (B).
			\item Feature branches with daily commits encouraged (C).
		\end{itemize} &
		\begin{itemize}[leftmargin=*]
			\item Reduces merge conflicts (A, B, C).
			\item Rapid user feedback (B).
		\end{itemize} &
		\begin{itemize}[leftmargin=*]
			\item A developer works in isolation until merge (C).
		\end{itemize} \\
		   	\midrule
		% P5: Every commit should build the mainline on an integration machine
		\rowprcheader{Every commit should build the mainline on an integration machine}{practice perception} \\
		\begin{itemize}[leftmargin=*]
			\item Builds on PR and mainline levels (A, B, C).
		\end{itemize} &
		\begin{itemize}[leftmargin=*]
			\item Ensures test execution (A).
			\item Faster feedback (B).
			\item Minimizes regression bugs (C).
		\end{itemize} &
		\begin{itemize}[leftmargin=*]
			\item Bottleneck due to PR review (A).
			\item Frequent PRs swamp reviewers (A).
			\item Build infrastructure is bottleneck (C).
		\end{itemize} \\
		   	\midrule
		% P6: Keep the build fast
		\rowprcheader{Keep the build fast}{project context and practice perception} \\
		\begin{itemize}[leftmargin=*]
			\item One minute to a few hours (A).
			\item Ten minutes or less (B).
			\item Ten to twenty minutes (C).
		\end{itemize} &
		\begin{itemize}[leftmargin=*]
			\item Less developer context switching (A, B).
			\item Reduces developer blocking (B).
		\end{itemize} &
		\begin{itemize}[leftmargin=*]
			\item Reliance on external services in build is time-consuming (A).
			\item Infrastructure runs tests inefficiently (C).
			\item Decreases perceived velocity (C).
		\end{itemize} \\
		   	\midrule
		% P7: Test in a clone of the production environment
		\rowprcheader{Test in a clone of the production environment}{practice perception} \\
		\begin{itemize}[leftmargin=*]
			\item Environment flags in the Dockerfile (A, B, C).
		\end{itemize} &
		\begin{itemize}[leftmargin=*]
			\item Prod and dev environments in sync, consistent tests (A).
			\item Ensures consistent tests across builds (B).
			\item Application execution consistency (C).
		\end{itemize} &
		\begin{itemize}[leftmargin=*]
			\item Scalability issues due to resource restrictions on dev environments (A).
			\item More maintenance (B).
			\item Complete clones hard due to non-transferable aspects (B, C).
		\end{itemize} \\
		   	\midrule
		% P8: Make it easy for anyone to get the latest executable
		\rowprcheader{Make it easy for anyone to get the latest executable}{practice perception} \\
		\begin{itemize}[leftmargin=*]
			\item Docker facilitates project spin-up (A, B, C).
		\end{itemize} &
		\begin{itemize}[leftmargin=*]
			\item Reduced build complexity (A, C).
			\item Easier onboarding (B).
		\end{itemize} &
		\begin{itemize}[leftmargin=*]
			\item Reducing build complexity requires effort (B).
		\end{itemize} \\
			\midrule
		% P9: Ensure that system state and changes are visible:
		\rowprcheader{Ensure that system state and changes are visible}{} \\
		\begin{itemize}[leftmargin=*]
			\item Dedicated Slack channel (A, B, C).
		\end{itemize} &
		\begin{itemize}[leftmargin=*]
			\item Keeps everyone in the loop (A, B, C).
		\end{itemize} &
		\begin{itemize}[leftmargin=*]
			\item Notification fatigue from frequent builds (A).
			\item Requires discipline and collides with responsibilities (B, C).
		\end{itemize} \\
			\midrule
		% P10: Automate deployment
		\rowprcheader{Automate deployment}{practice perception} \\
		\begin{itemize}[leftmargin=*]
			\item Via manual trigger. Not all can trigger for specific projects (A).
			\item Via manual trigger. Everyone but newest members can trigger (B).
			\item Via manual trigger once a day, except on weekends. Not everyone deploys (C).
		\end{itemize} &
		\begin{itemize}[leftmargin=*]
			\item Minimizes deployment effort (A).
			\item Maintains security (A, C).
			\item Consistency and repeatability (B).
			\item Everyone deploying creates ownership culture (B).
			\item Minimizes deploying breaking changes over period of non-activity (C).
		\end{itemize} &
		\begin{itemize}[leftmargin=*]
			\item No testing in deployment builds to reduce build time (B).
			\item Requires preparation (C).
		\end{itemize} \\
		\bottomrule[1pt]
	\end{tabularx}
\end{table*}

\subsection{Case Profiles}
\label{subsec:profiles}
%
%Each of the three organizations we investigated have different use cases and rationale for applying continuous practices.
%In the following subsections, we describe that context for ORGs A, B, and C.

All three organizations employ the pull request (PR) model \cite{gousios2015work}: a developer makes their change locally, then creates a pull request which is approved by other members of the team and then merged into the mainline.
At various stages of that workflow, automated builds and tests can occur: local builds and unit tests, builds and associated PR tests on the pull request pre-merge, builds and tests (regression, integration) on the mainline after merging, and deployment builds and acceptance tests once a release is ready.

\subsubsection{Organization A}

Organization A is involved in the data business, which includes collecting and processing large amounts of data from users around the world.
They use an assortment of continuous tools as part of their operations.
In particular, the CI tool Jenkins plays a crucial role in their software development, testing, and deployment. 
Organization A has adhered to continuous practices since it was founded.
As data is paramount to organization A, several teams and corresponding projects are dedicated to processing and collecting data.
These projects include reporting, data processing pipelines, and data collection pipelines.
However, organization A's business and operations are multi-faceted. 
There are also projects that include user-facing content management websites.

Organization A adopted continuous practices when the organization was founded due to the perceived benefits of consistency and reproducibility offered by CI.
Given CI's reputation for reducing bugs, particularly integration problems, CI presents organization A with the opportunity to ensure the robust flow of their data, from collection to processing.
Organization A saw that using automation and CI tools would help ensure that the organization's changes are not tested in snowflake environments, as environments would be kept in sync by Docker.
Moreover, organization A perceives that challenges with onboarding are alleviated due to the ease of using Docker to help set up and run development environments on developer machines.

\subsubsection{Organization B}

Organization B provides online content through their publishing platform, including advertisement management.
They have invested a considerable amount of effort into turning their original stand-alone product into a platform that can easily accommodate new components and their supporting infrastructures.
They consider themselves continuous and are committed to furthering their continuous practices.
Among the three organizations, organization B maintains the distinction of having a highly structured and automated workflow.

The primary reason for adopting continuous practices was to receive rapid feedback on changes made to their product and the efficiency provided by automation.
The desire for rapid feedback was twofold.
First, it cultivated the feeling among developers that the changes they were making had a quick, observable impact on the overall platform they deploy. 
Second, to management it was more about following agile practices and eliciting feedback from users as often as possible.
Automating their process would enforce a strict workflow and remove manual processes, allowing effort to be focused on higher priority issues.
Not only is their development team deeply embedded in continuous practices, but almost the entire organization continuously monitors how the platform is performing with respect to tracking real-time revenue.

\subsubsection{Organization C}

Organization C provides an online booking platform that integrates with multiple existing services and caters to many customers worldwide.
They report themselves as being \emph{almost continuous} for over a year as they attempt to integrate the practices into their workflow.
However, based on participant responses, it seems that their perception of continuous practices focuses on practices 2, 3, 5, and 10: build and deployment automation, testing, and triggering mainline builds.
Jenkins and Docker are the tools of choice when it comes to build automation, with a migration to Kubernetes underway to facilitate infrastructure management and increase product reliability since their product is expected to fulfill service-level agreements.

Organization C decided to adopt continuous practices because they felt such practices would accommodate their larger, more complex product, enhance its quality, and improve the development workflow.
Organization C is structured around their online booking platform, with dedicated teams supporting its various aspects.
The development team is responsible for adding new features, maintaining the product, and fixing bugs.
The configuration team focuses on creating separate product configurations for each customer.
The support team deals with inbound customer requests and, if necessary, forwards any unresolvable issues to the development team.

\subsection{Differences in How and Why Companies Use CI}
\label{subsec:differences}
In this section, we focus on the differences in how the three organizations use CI and the rationale they provided.
We focus only on the CI practices where the companies diverge greatly and where the differences highlight important considerations. 
Table \ref{tbl:practices_orgs} briefly covers all ten practices, including rationale, benefits, and challenges. Finally, even more detailed, rich descriptions of \emph{all practices per organization} can be found in the organization profiles provided in our reproduction package.

% Differences in repo structure
\subsubsection{Maintain a single source repository}
This practice is supposed to make builds easier, enhance code visibility, and centralize dependencies (from Table \ref{tbl:practices_lit}). 

Each organization applies this practice in a different way.
Because organization A's priority is to reduce merge conflicts, project components are often separated based on which components need simultaneous updates, as P3 mentioned: \quote{Because all the three pieces just interact with one another, they're not sort of set up really to like produce something for another project.}
However, the logical boundaries between the projects are not always clear.

On the other hand, organization B opted for a microservice-like project separation with each component of their platform isolated in its own repository.
When used as a dependency for another project, a component will be fetched from its repository or a dependency management system and be included in a project's build process.
Organization B did not consider using a single source repository, mainly because the advantages it offered over their current repository organization were not clear, as P9 put it: \quote{No one's been able to show us the benefit of mono repo or I have not been able to see the benefit of a mono repo.}
As a result, organization B developers often find themselves having to duplicate their workflow and tooling across components, and having to maintain complex dependency graphs.

Finally, in the case of organization C, because their entire workflow is built around a single project (their online booking platform), that project exists in a single repository.
Over time, new components that were developed in-house for the booking platform were separated into their own repositories, with older components that provided tooling and other cross-cutting concerns remaining part of the main project's repository.
The reason for this repository organization was a tool limitation, as P12 noted: \quote{The biggest reason for that was file storage, like git LFS wasn't mature enough at the time.}

\vspace{-1mm}
\begin{mdframed}
\textbf{Finding:} Differences in repository structures across organizations are due to practice perceptions (dealing with merge conflicts had a higher priority than the unclear benefit of using a single source repository).
\end{mdframed}
\vspace{-1mm}

% Differences in testing
\subsubsection{Make the build self-testing}
All three organizations run unit tests in their pull request builds and additional regression tests in their mainline builds.
Organization A prioritizes faster builds pre-PR, with lengthier tests performed by reviewers: \quote{I don't want long tests on the developer's side (P1).}

Organization B prioritizes financial resources for their production infrastructure over their development infrastructure, and thus adds regression tests to the mainline build, but not to the pull request builds.
They no longer run end-to-end or smoke tests because those proved to be flaky in their execution, as P6 noted: \quote{We found that our end-to-end testing and smoke tests were much less useful because they had lots of false positives because they involved the real world}. 
%. The tests involved like running the site through headless Chrome which has to get assets from 3rd parties which change and become weird and sometimes those requests timeout.}
Furthermore, they sacrifice coverage for speed, opting to run less comprehensive tests to minimize the amount of time a developer spends waiting for a build.

Organization C performs the most comprehensive testing. Pull request builds contain several types of tests, including end-to-end tests.
They prioritize thorough testing over shorter build durations.
The same type of build is triggered on the mainline when a change is merged, then a dedicated quality assurance (QA) team performs manual tests.
Then once a feature is complete, a product manager tests the application again, as P13 described: \quote{There will be automation and manual sanity checks after it's ready to go.}
While this minimizes  breaking changes, it increases build duration and decreases development velocity.

All three organizations also report doing some form of manual UI testing.
For organizations A and B, this is due to a trade-off between the benefits automated UI testing brings vs. how much effort is required to set it up.
P1 described it as: \quote{It would be too much work. There wouldn't be enough value for how much money I have to spend to get that testing to automate all that selenium. All that'd be too cost prohibitive effectively, you know with how much websites change.}
For organization C, manual UI testing is part of their quality assurance process.

% Where is the workload situated (developers vs. automation) - We have that info
The human effort required to verify/review new changes before merging into mainline varies.
Organization B has the least effort required to verify a change with only the responsible developer doing the review process.
Although a change used to require two independent reviewers to sign off on the change, organization B wants to get rid of that practice, preferring to find any problems in deployment and production testing (e.g., with canary deploys \cite{rossi2016continuous}).
Organization C has the most manual effort required to verify a change with developers, QA personnel, and product managers reviewing changes throughout the development process.
Organization A is in the middle with both developers and a limited pool of reviewers (team leads) required to verify changes.

\vspace{-1mm}
\begin{mdframed}
\textbf{Finding:} Differences in testing are due to tool limitations (with automated UI tests not being feasible), practice perceptions (the perceived work required for integration tests outweighs the perceived benefits, and data-centric testing outweighs having shorter builds), and project context (data-centric tests require communication with external services).
\end{mdframed}
\vspace{-1mm}

% Differences in build durations
\subsubsection{Keep the build fast}
Organization A has builds ranging from one minute to a few hours depending on the project the build is attached to and its complexity.
Typical software application projects at organization A do not have lengthy builds, which helps reduce developer context switching because they do not need to switch tasks while waiting for a build to run.
However, because organization A's business model relies on user data, some projects involve extensive data processing or need to communicate with services external to the build pipeline.
This can increase build duration considerably, and can be problematic with failed builds, as P2 mentioned: \quote{Yeah, I guess the only issue would be if it ran for 40 minutes and then it failed and then you got to go in and figure out why it failed then re-push.}

Organization B's builds typically take 10 minutes or less for two reasons: reducing developer context-switching, and reducing developer blocking, as P8 put it: \quote{I don't really think about it or feel that it's holding me up}.
While these are the driving reasons behind this practice, it is possibly facilitated by their repository structure which separates components and tests them in isolation, thus each build only focuses on testing the project it is attached to.
Another factor that could facilitate faster builds is the less comprehensive testing they conduct within the builds.

Organization C's builds typically take between 10 and 20 minutes.
This is influenced by two factors: comprehensive testing being of a higher priority than build duration, and their infrastructure.
Due to infrastructure limitations, some of organization C's test suites run inefficiently in the build, as P12 put it: \quote{So we actually have some [test] serialization inside of each child as well.}
This results in a decrease in their perceived development velocity.

\vspace{-1mm}
\begin{mdframed}
\textbf{Finding:} Differences in build times are due to project context (data-centric projects have longer builds) and practice perceptions (doing comprehensive testing outweighs having shorter builds).
\end{mdframed}
\vspace{-1mm}

% Differences in deployment
\subsubsection{Automate deployment}
All three organizations require the deployment process to be triggered manually.
The main differences are in who is allowed to deploy changes to production and when this happens.
For organizations A and C, deployment is usually a privilege assigned to a senior member of the development team.
This helps \textbf{maintain security} and ensures no unapproved changes make their way to production, as P1 noted: \quote{Just so not anybody can just do it whenever they want as a team grows.}
However, anyone can ask for deployment permissions.
For organization B, deployment is the responsibility of the developer who authored the change they want to deploy.
This helps foster a culture of ownership, as P7 put it: \quote{If it's their code, we usually tell them to kind of babysit it all the way through because when deploying you need to know the context, you know what to look for when it's actually out and deployed. And the person who knows best is the person who wrote it. So as much as possible, we like to have that person press the button.}
This helps foster \textbf{a culture of ownership} where the developer \emph{owns} the change they made all the way to production.

While deployments for organizations A and B are more frequent, organization C has a stricter deployment schedule.
Due to the amount of testing that happens and the pre-deployment preparation required (testing and writing up release notes), deployments are scheduled once a day.
In these deployments, a collection of features is pushed to production and made available to customers.
While such deployments happen daily, they do not occur before weekends or long periods of inactivity to avoid pushing breaking changes when no one is around to fix them, as P11 mentioned: \quote{Well if we deploy things on Fridays, it's when we're winding down for the week. And if we encounter any unexpected behavior, especially maybe our European customers encounter unexpected behavior, that means they're reaching out to us at 2 in the morning on Friday night.	I don't have a team in place to be able to handle those requests.}

\vspace{-1mm}
\begin{mdframed}
\textbf{Finding:} Differences in deployment procedures are due to practice perceptions (enhancing security by enforcing deployment privileges for some organizations outweighs a sense of change ownership).
\end{mdframed}
\vspace{-1mm}

\section{Discussion}
\label{sec:discussion}

Our findings show that all ten core CI practices are implemented at our three case study companies to some extent, but \textbf{how} they are implemented varies (see Table \ref{tbl:practices_orgs}). Three themes help explain this variation: differences in project context; differences in how CI practices are perceived; and constraints that negatively impact fast delivery of software (one of the original motivations for CI in general).

\subsection{Differences Related to Project Context}
\label{subsec:context}
The way CI practices are implemented depends on the contextual factors of the project they are intended to support. These contextual factors in the three organizations we studied include differences in project type, testing strategy, and CI infrastructure.

% Project type
The \emph{type of project} influences how organizations prioritize and value the different practices.
Organization A, for example, has a heavily data-centric application and tolerates longer build and test cycles because their data-centric tests require communication with external services and fetch large volumes of data. This trade-off, where builds might take up to two hours, is essential to maintain confidence in the data processing components.
Organization B, on the other hand, values faster build times in order to minimize feedback cycle times.

%Testing Strategy
\emph{Testing strategy} is another important contextual factor. Part of the testing strategy is the relative weight a team places on unit tests, integration tests, and regression tests. If a particular test strategy did not add value---for example, if integration testing did not provide actionable and informative results---the three organizations did not prioritize it.
For example, one team in organization A focuses on unit tests due to their low cost of implementation compared to other forms of testing> They also use manual testing in place of integration tests, because as P1 put it: \quote{[Writing integration tests] would be too much work.}
Organization C includes many different types of tests in their builds because of the high priority placed on ensuring that changes do not break a build.

Another difference we found relates to usability testing.
Developers and other team members are required to integrate manual testing into their workflows to ensure their changes work.
For organizations A and B, this is done by developers, but organization C has an additional gate of Q/A personnel dedicated to interface testing.

This difference in testing strategy leads to organizational differences in build speeds, deployment procedures, and the self-testing nature of the build.

The final contextual factor we identified concerns \emph{differences in CI infrastructure}.
Build farms and CI workers are costly but necessary to reduce slow build queues \cite{gallaba2019improving} or slow test case execution \cite{elbaum2014techniques}.
For the three organizations, the differences in their infrastructure impacts their implementation of CI practices.
Both organizations A and B use an AWS infrastructure rather than hosting locally as organization C does. This makes faster builds possible for A and B. In organization A's case, testing would not even be possible in a locally hosted CI configuration.
On the other hand, organization B's ability to duplicate their production infrastructure for testing purposes is impacted by cost constraints, which leads to compromising how many instances are run in a testing environment versus a production environment, as P8 stated: \quote{The only thing that would be different between them is the size and the number of instances of things.}
Organization A does not have similar strict cost constraints.

Organizational context is a major factor in understanding what it means for an organization to call itself ``continuous''. For example, if queried about their implementation of the ``make the build self-testing'' practice, all three might state that they follow this practice, although as we have shown the actual implementation of this practice can vary widely (from fully self-testing with no human automation, to significant manual intervention in the testing process).

\textbf{Implications for researchers:} Projects have unique tooling and scaffolding as well as their own unique requirements based on their intended functionality.
When investigating and drawing comparisons about CI practices across projects, one must control for these confounding factors.

\textbf{Implications for practitioners:} Practices and tooling facilitate project development.
When implementing practices or adopting tools to enhance a development workflow, one must prioritize those that serve the project's workflow and intended functionality, and recognize the constraints tools and practices may bring.
There is no \emph{single way} of implementing a particular CI practice.

\subsection{Differences Related to Practice Perception}
\label{subsec:perception}
The second dimension along which we observed differences in CI practice implementation was that of \emph{CI practice perception}.
The term \emph{practice perception} refers to an organization's interpretation of how a CI practice should be implemented and its perceived benefits.
The way the organizations perceived a practice led to significant differences in its implementation. We look at two practices as examples.	

% single source practice
The CI practice ``maintain a single source repository'' was followed differently at all three organizations. 
This practice refers to the ability of a project at an organization to be built without requiring build artifacts (code, libraries, etc.) from other projects. 
All three organizations claimed to follow this practice. 
However, the implementations of the practice varied widely.

Within organization A, some projects use multiple repositories. 
Components such as backend and frontend services are separated into individual component repositories, but deployment treats these components as a single application.
This is because organization A found the more separated their components are, the fewer merge conflicts arise when developers work in isolation on these independent components.

Organization B developers prefer to keep their dependencies separate from the deployable projects, resulting in a microservice-like architecture during the build process, further compounding build complexity.

Organization C developers started out with a single source repository, but have since moved to a hybrid between single source repository and microservice models where some dependencies live within the main project's repository and others are fetched at build time.

Even though all three organizations claimed to follow the CI practice of maintaining a single source repository, wide implementation differences exist: organization A prioritized ease of integration, organization B preferred high cohesion and low coupling, and organization C was restricted by git limitations at the time.

Implementation differences also exist for the practice of ``automate deployment". Automating deployment should speed up release cycles and reduce errors due to manual steps, and again all three organizations claimed to follow this practice. 

However, a team lead at organizations A and C has to push the deployment ``button''. At organization B, the individual developer presses this button. For organizations A and C, this responsibility is manual to prevent potential bad actors deploying to production.
Organization B leaves this responsibility with individual developers and relies on deployment feedback.
Automation also varied by \emph{when} deployment happened. At C, deployment is more structured, and follows a well-defined once-daily cycle. At A and B, deployment happened whenever a feature was finished.

% summary
An organization's perceptions of a CI practice affected how it was implemented and what value the organization thought it would bring.
But can all these widely varying implementations bring the same value? 
Many practitioner-focused CI guidelines \cite{humble2010continuous,forsgren2019accelerate} highly recommend the use of concrete measures to assess the value of CI practices. 
For example, measuring cycle time or the time to implement a feature provides concrete information on the effectiveness of development workflows.

However, none of the three organizations had implemented these measures. 
Organization B relied on their \textit{revenue} to measure the impact of their practices, while organizations A and C relied on subjective assessment via customer response instead. 
But revenue and customer response have other inputs that make the value of CI practices hard to perceive.

\textbf{Implications for researchers:} Bad CI practices or anti-patterns \cite{duvall2011continuous, zampetti2020empirical, gallaba2018use} assume that a correct version of a practice or configuration exists and is desired.
However, when taking into account what benefits practitioners associate with the individual CI practices, this is not always desired.
It is often sufficient for an organization's implementation of a practice to be \emph{good enough} rather than correct.
One should control for such perceived benefits when investigating the correctness of a practice implementation.

\textbf{Implications for practitioners:} Concretely monitoring a project's workflow to determine practice value is important when considering implementing new practices or modifying existing ones.
Teams should consider the metrics they wish to improve to determine which practices are best suited to the task.
Similarly to the findings we report in Section \ref{subsec:context}, there is no \emph{single way} of implementing a CI practice.

\subsection{CI-Related Process Constraints}
\label{subsec:constraints}
% (Industry)
% 	> Tool-induced bottlenecks
% 		>> The impact of infrastructure constraints on build speed and test execution
% 		>> The mainline build becomes a single point of failure at ORG B
%	> Process-induced bottlenecks
% 		>> The reviewer paradigm at ORG A (with which ORG B is doing away)
% 		>> Too many testing gates for ORG C
A major reason for using the ten core CI practices is to reduce software development feedback cycle time (i.e., the time from a feature being developed to when customer feedback is received) \cite{fowler2000continuous}.
However, introducing CI constrained other aspects of the software development workflow at each company, increasing cycle time.

% , having moved to CI you now find problems in how quickly you can get software out.
Implementing a CI pipeline and associated automation introduced bottlenecks in the flow of work.
For example, organization B developers are responsible for merging their own changes, and new PRs must wait until other PRs are finished being merged.
When the initial PR is an important change, developers working on other PRs have to wait for all build failures and CI process checks to pass.
This slows cycle time, blocks developers, and causes context-switching, the opposite of the intended benefit.

One solution we came across in our interviews was the use of feature-flags, as stated by P14: \quote{I've been a big proponent of feature flags ... you can develop as much as you want.} 
However, feature flags have been shown to introduce issues of their own, such as technical debt and maintenance overhead \cite{rahman2016feature}.

The second type of constraint is in development and testing workflows.
Organization C's extensive testing approach, one which requires manual testing for user-facing changes, happens on several levels, as described in Section \ref{subsec:differences}.
The many testing gates compromise feedback cycle time.
On the development side, when coupled with smaller, more frequent code contributions---particularly those that do not constitute a cohesive, testable feature---fast, automated delivery can diminish the ability to manually test features upon their release.
This is contrary to what was proposed by Rahman et al. \cite{rahman2018characterizing}.
For example, product managers and other non-technical team members are required to manually review implemented features to make sure requirements are fulfilled.
However, they have trouble mapping code changes to features.

\textbf{Implications for researchers:} CI practices have several benefits, but they also have limitations.
While the ``mining'' style of investigation is perfect for drawing conclusions about the effectiveness of automation or the CI methodology as a whole, it may not be the most powerful at identifying the benefits and challenges of using CI.
In fact, it may add threats to the way CI data is analyzed and interpreted.

\textbf{Implications for practitioners:} CI is not a \emph{silver bullet}.
It has limitations that could hinder development workflow depending on how its individual practices are implemented.

\section{Threats to Validity}
\label{sec:threats}
% Total Quality Framework Intro
In this section, we discuss the limitations of our study and how we attempted to mitigate them.
Using the total quality framework by Roller and Lavrakas \cite{roller2015applied}, we break this section down based on credibility, analyzability, transparency, and usefulness.

\subsection{Credibility}
Credibility is related to the accuracy and completeness of the data collected.
This deals with two aspects: scope and data gathering.

\subsubsection{Scope}
% External validity
The scope of our study was limited to our participants, three software development organizations that provide software as a service (SaaS) in our immediate geographic area.
Furthermore, due to the opportunistic nature of the interviewee recruitment within each organization, our observations may not be representative, though we mitigated that by interviewing a cross-section of organizational roles.
% These organizations were recruited opportunistically, and as such our findings may not generalize well.
%It is possible that software development organizations with a different business model or contexts will face different challenges.
%However, since we focus on CI practices, with further research it is possible that our findings generalize across different contexts.

\subsubsection{Data Gathering}
% Construct validity
% Inter-researcher reliability
% Internal validity
With respect to our constructs, and consequently our data gathering, we followed the same approach as Zhao et al. \cite{zhao2017impact} and M{\aa}rtensson et al. \cite{maartensson2017continuous} where they operationalized practices for particular project contexts.
However, we did this across organizational contexts.
To further ensure the practices we selected were rooted in reality, we used those in the original blog post by Fowler and Foemmel \cite{fowler2006continuous} and later elaborated on by Humble and Farley \cite{humble2010continuous}.

To mitigate researcher bias, which is typically introduced when a single researcher codes a significant amount of collected data, we did the following:
\begin{itemize}
	\item \textbf{For interviews:}
	There were always two interviewers present at each interview.
	We followed a list of prepared questions, asking for elaboration where necessary, and made sure to ask participants to repeat or rephrase unclear sections of their responses as a form of member-checking.
	\item \textbf{For transcript coding:}
	We coded the transcripts in pairs.
	Following the approach by Heikkl{\"a} et al. \cite{heikkila2017managing}, once a transcript was coded, the two coders would meet in what was referred to as an ``agreement session''. This was done in order to discuss differences in coding schemes and settle on a solution that aligned with both coders' interpretation of the data.
	\item \textbf{For thematic analysis:}
	Our thematic analysis was conducted in a similar fashion.
	The agreed upon themes were then included in a member-checking survey that was sent to the interviewees to measure agreement.
\end{itemize}
Afterwards, copies of the case reports pertaining to each organization were sent back to that organization to make sure our understanding was accurate.
Finally, a copy of this paper was sent to the participating organizations pre-submission to ensure that we did not misrepresent or misinterpret interviewee responses.

\subsection{Analyzability}
Analyzability focuses on the accuracy of the analysis conducted and the interpretations drawn from the study.

\subsubsection{Processing}
% Processing (transcriptions)
The interviews were recorded after obtaining the interviewees' consent, and were transcribed using a voice-to-text transcription tool.
These transcripts were later verified by listening to the actual interviews and comparing them against the transcripts to minimize errors.

\subsubsection{Verification}
% Verification (triangulation between participants)
We triangulated by interviewing a cross-section of organization members, ensuring our information was not biased by a single perspective.
We constructed a member-checking survey that included the themes that emerged from the interviews and sent that to our interviewees.
We received 8 out of 18 responses (44.4\%).
Furthermore, we sent the case reports we constructed as well as this paper to our contacts in the three organizations to enhance our verification process.
We received positive responses from all three organizations, along with minor fixes as discussed in section \ref{sec:study}.
Where possible, we bolstered our analysis by mining the organizations' development activity logs.
These logs, however, only provided verification for certain practices (2, 4, 5, and 6) and are included in the reproduction package.

\subsection{Transparency}
% Reporting (providing quotes, rich details, transferability to other contexts)
% Mention NDA here
We attempted to provide rich details and quotes wherever possible, as well as a reproduction package.
However, we were bound by non-disclosure agreements regarding certain aspects of the interviewees' responses.
As such, we are unable to include the transcripts of our interviews, only our interview questions, codebook, thematic analysis tables, case reports, and anonymized activity logs.

\subsection{Usefulness}
% How useful are our results given the existing limitations?
Usefulness indicates how actionable the results from this study are, and how well they transfer to other contexts (external validity).
Our results are organization- and workflow-centric, and based on ten widely accepted CI practices.
% We argue that they are applicable to most development teams who work on SaaS products in small and medium-sized organizations.
% They may be applicable to other organizations so long as their team structure is comparable, and the nature of their product is similar to that of an online service.
However, we are making the point that context must be considered when investigating these practices and, as such, organizations similar to ones we studied constitute the best candidates for finding transferability.
Our approach to understanding and analyzing the use of these practices is transferable to other organizations that are considering adopting CI practices because project context is important and it influences how people perceive these practices.

For example, if a development team in a company that has different characteristics (such as number of employees, or different end-product) decided to adopt CI practices, they would still need to consider them along the dimensions we have identified: project context and existing tooling, process constraints, and perception.
Consequently, setting their operational thresholds will depend on their particular context, constraints, and perception of CI practices.
We include implications both for practitioners and researchers that open up several avenues of research.
\section{Conclusion and Future Work}
\label{sec:conclusion}

Our multi-case study of three software-as-a-service organizations helped us determine the extent of CI practice implementation based on the original ten practices proposed by Fowler and Foemmel \cite{fowler2006continuous}.
We reported \textbf{how} these CI practices were adopted by the organizations and how they delivered the \textbf{benefits} claimed in the literature. 
We also identified the \textbf{challenges} that may result from implementing these practices.
How the organizations implemented the ten continuous practices depended on three things: the organizational perception of these practices (Section \ref{subsec:perception}), the project context within which they are applied (Section \ref{subsec:context}), and CI-related process and tool constraints (Section \ref{subsec:constraints}).

In Section \ref{subsec:differences}, we highlighted how context, perception, and constraints influence CI practice implementation.
For example, how organizations do or do not maintain a single repository depends on the organizational context.
Although organizations may want to minimize merge conflicts, there are other forces that influence their decision in this regard (such as the need for an organization to do updates simultaneously on different project components that may have dependencies between them).
How each organization conducts testing also varies depending on testing impact on build times and a need to do manual or data-centric testing.
Similarly, keeping the build fast may seem like an ideal goal, but again it isn't always possible (e.g., data-centric projects have longer builds). 
Finally, automated deployment may be influenced by security needs and deployment privileges because organizations perceive security to be more important than frequent deployment.

The difference in practice implementation raises questions to researchers as to whether CI should continue to be investigated as a single practice (which it is not), and thus if its claimed effects when studied can be attributed to CI as a whole or to perhaps one or more individual practices.
What we found through our study is that although organizations and researchers tend to consider CI as one thing (either a company is or is not continuous), we found that how CI is adopted varies significantly across even three organizations that follow a similar software business model (see Table \ref{tbl:practices_orgs}).

We also showed how understanding the different pressures and nuances of CI and its benefits can only be understood using qualitative methods, such as interviews.
Understanding why CI practices are not adopted cannot be determined from data alone (as there is no trace to follow).
Thus, analysis of CI data only shows part of the picture. 
We recognize, however, that our findings are limited to three organizations and that more studies should be conducted.

In future work, we aim to develop a theory of CI which captures the practices and project context factors as the main constructs, and the benefits and challenges those practices may lead to as additional constructs. 
Understanding the trade-offs among practices in particular contexts is another future research avenue.

A theory about CI could be used by researchers to test specific hypotheses about practice use and the anticipated benefits/challenges of practice adoption, while practitioners can use the theory to help guide them how to use specific CI practices.
A theory may also provide them with guidance on which practices to not to adopt, which challenges to anticipate, and how their project context and the types of tools they use may impact those decisions. 
Currently, companies and researchers are somewhat in the dark when they think about CI and have not fully engaged with what we show is a very complex concept.

\ifCLASSOPTIONcompsoc
  \section*{Acknowledgments}
\else
  \section*{Acknowledgment}
\fi

We acknowledge the support of the Natural Sciences and Engineering Research Council of Canada (NSERC).
We would also like to thank Cassandra Petrachenko for her valuable help with this study.

\ifCLASSOPTIONcaptionsoff
  \newpage
\fi

\bibliographystyle{IEEEtran}
\bibliography{main}

\begin{thebibliography}{10}
\providecommand{\url}[1]{#1}
\csname url@samestyle\endcsname
\providecommand{\newblock}{\relax}
\providecommand{\bibinfo}[2]{#2}
\providecommand{\BIBentrySTDinterwordspacing}{\spaceskip=0pt\relax}
\providecommand{\BIBentryALTinterwordstretchfactor}{4}
\providecommand{\BIBentryALTinterwordspacing}{\spaceskip=\fontdimen2\font plus
\BIBentryALTinterwordstretchfactor\fontdimen3\font minus
  \fontdimen4\font\relax}
\providecommand{\BIBforeignlanguage}[2]{{%
\expandafter\ifx\csname l@#1\endcsname\relax
\typeout{** WARNING: IEEEtran.bst: No hyphenation pattern has been}%
\typeout{** loaded for the language `#1'. Using the pattern for}%
\typeout{** the default language instead.}%
\else
\language=\csname l@#1\endcsname
\fi
#2}}
\providecommand{\BIBdecl}{\relax}
\BIBdecl

\bibitem{fowler2000continuous}
M.~Fowler and M.~Foemmel, ``Continuous integration,''
  \url{https://tinyurl.com/ycbl2uhj}, 2006, [Online; accessed 21-May-2020].

\bibitem{vasilescu2015quality}
B.~Vasilescu, Y.~Yu, H.~Wang, P.~Devanbu, and V.~Filkov, ``Quality and
  productivity outcomes relating to continuous integration in github,'' in
  \emph{Proceedings of the 2015 10th Joint Meeting on Foundations of Software
  Engineering}, 2015, pp. 805--816.

\bibitem{staahl2019big}
D.~St{\aa}hl, A.~Martini, and T.~M{\aa}rtensson, ``Big bangs and small pops: on
  critical cyclomatic complexity and developer integration behavior,'' in
  \emph{2019 IEEE/ACM 41st International Conference on Software Engineering:
  Software Engineering in Practice (ICSE-SEIP)}.\hskip 1em plus 0.5em minus
  0.4em\relax IEEE, 2019, pp. 81--90.

\bibitem{rossi2016continuous}
C.~Rossi, E.~Shibley, S.~Su, K.~Beck, T.~Savor, and M.~Stumm, ``Continuous
  deployment of mobile software at facebook (showcase),'' in \emph{Proceedings
  of the 2016 24th ACM SIGSOFT International Symposium on Foundations of
  Software Engineering}, 2016, pp. 12--23.

\bibitem{hilton2016usage}
M.~Hilton, T.~Tunnell, K.~Huang, D.~Marinov, and D.~Dig, ``Usage, costs, and
  benefits of continuous integration in open-source projects,'' in \emph{2016
  31st IEEE/ACM International Conference on Automated Software Engineering
  (ASE)}.\hskip 1em plus 0.5em minus 0.4em\relax IEEE, 2016, pp. 426--437.

\bibitem{fowler2006continuous}
M.~Fowler and M.~Foemmel, ``Continuous integration,''
  \url{https://tinyurl.com/y8d3asjv}, 2006, [Online; accessed 21-May-2020].

\bibitem{fitzgerald2017continuous}
B.~Fitzgerald and K.-J. Stol, ``Continuous software engineering: A roadmap and
  agenda,'' \emph{Journal of Systems and Software}, vol. 123, pp. 176--189,
  2017.

\bibitem{guerriero2019adoption}
M.~{Guerriero}, M.~{Garriga}, D.~A. {Tamburri}, and F.~{Palomba}, ``Adoption,
  support, and challenges of infrastructure-as-code: Insights from industry,''
  in \emph{2019 IEEE International Conference on Software Maintenance and
  Evolution (ICSME)}, 2019, pp. 580--589.

\bibitem{shahin2017continuous}
M.~Shahin, M.~A. Babar, and L.~Zhu, ``Continuous integration, delivery and
  deployment: a systematic review on approaches, tools, challenges and
  practices,'' \emph{IEEE Access}, vol.~5, pp. 3909--3943, 2017.

\bibitem{maartensson2017continuous}
T.~M{\aa}rtensson, P.~Hammarstr{\"o}m, and J.~Bosch, ``Continuous integration
  is not about build systems,'' in \emph{2017 43rd Euromicro Conference on
  Software Engineering and Advanced Applications (SEAA)}.\hskip 1em plus 0.5em
  minus 0.4em\relax IEEE, 2017, pp. 1--9.

\bibitem{humble2010continuous}
J.~Humble and D.~Farley, \emph{Continuous Delivery: Reliable Software Releases
  through Build, Test, and Deployment Automation}.\hskip 1em plus 0.5em minus
  0.4em\relax Pearson Education, 2010.

\bibitem{laukkanen2017problems}
E.~Laukkanen, J.~Itkonen, and C.~Lassenius, ``Problems, causes and solutions
  when adopting continuous delivery—a systematic literature review,''
  \emph{Information and Software Technology}, vol.~82, pp. 55--79, 2017.

\bibitem{berg2019common}
K.~Berg, ``4 common problems with continuous integration and deployment and how
  to avoid them,'' \url{https://tinyurl.com/ybp9n9cn}, 2019, [Online; accessed
  14-July-2020].

\bibitem{heller2015continuous}
M.~Heller, ``Continuous integration: The answer to life, the universe, and
  everything?'' \url{https://tinyurl.com/y8rhxnt8}, 2015, [Online; accessed
  14-July-2020].

\bibitem{bugayenko2014continuous}
Y.~Bugayenko, ``Continuous integration is dead,''
  \url{https://tinyurl.com/ybsuvhxr}, 2014, [Online; accessed 14-July-2020].

\bibitem{jaspan2018advantages}
C.~Jaspan, M.~Jorde, A.~Knight, C.~Sadowski, E.~K. Smith, C.~Winter, and
  E.~Murphy-Hill, ``Advantages and disadvantages of a monolithic repository: a
  case study at google,'' in \emph{Proceedings of the 40th International
  Conference on Software Engineering: Software Engineering in Practice}, 2018,
  pp. 225--234.

\bibitem{zhao2017impact}
Y.~Zhao, A.~Serebrenik, Y.~Zhou, V.~Filkov, and B.~Vasilescu, ``The impact of
  continuous integration on other software development practices: a large-scale
  empirical study,'' in \emph{2017 32nd IEEE/ACM International Conference on
  Automated Software Engineering (ASE)}.\hskip 1em plus 0.5em minus 0.4em\relax
  IEEE, 2017, pp. 60--71.

\bibitem{duvall2011continuous}
P.~Duvall, ``Continuous delivery patterns and antipatterns in the software
  lifecycle,'' \url{https://tinyurl.com/yb2o78m6}, 2011, [Online; accessed
  07-July-2020.

\bibitem{staahl2013experienced}
D.~St{\aa}hl and J.~Bosch, ``Experienced benefits of continuous integration in
  industry software product development: A case study,'' in \emph{The 12th
  IASTED International Conference on Software Engineering}, 2013, pp. 736--743.

\bibitem{hukkanen2015adopting}
L.~Hukkanen \emph{et~al.}, ``Adopting continuous integration-a case study,''
  2015.

\bibitem{kikas2017structure}
R.~Kikas, G.~Gousios, M.~Dumas, and D.~Pfahl, ``Structure and evolution of
  package dependency networks,'' in \emph{2017 IEEE/ACM 14th International
  Conference on Mining Software Repositories (MSR)}.\hskip 1em plus 0.5em minus
  0.4em\relax IEEE, 2017, pp. 102--112.

\bibitem{zampetti2020empirical}
F.~Zampetti, C.~Vassallo, S.~Panichella, G.~Canfora, H.~Gall, and M.~Di~Penta,
  ``An empirical characterization of bad practices in continuous integration,''
  \emph{Empirical Software Engineering}, pp. 1--41, 2020.

\bibitem{islam2017insights}
M.~R. Islam and M.~F. Zibran, ``Insights into continuous integration build
  failures,'' in \emph{2017 IEEE/ACM 14th International Conference on Mining
  Software Repositories (MSR)}.\hskip 1em plus 0.5em minus 0.4em\relax IEEE,
  2017, pp. 467--470.

\bibitem{yin2003case}
R.~K. Yin, \emph{Case Study Research: Design and Methods}.\hskip 1em plus 0.5em
  minus 0.4em\relax London, U.K.: Sage Publications Inc., 2003.

\bibitem{storey2020software}
M.-A. Storey, N.~A. Ernst, C.~Williams, and E.~Kalliamvakou, ``The who, what,
  how of software engineering research: a socio-technical framework,''
  \emph{Empirical Software Engineering}, vol.~25, no.~5, pp. 4097--4129, 2020.

\bibitem{costello2019gartner}
K.~Costello, ``Gartner forecasts worldwide public cloud revenue to grow 17.5
  percent in 2019,'' \url{https://tinyurl.com/yalom8fc}, 2019, [Online;
  accessed 07-July-2020].

\bibitem{heikkila2017managing}
V.~T. Heikkil{\"a}, M.~Paasivaara, C.~Lasssenius, D.~Damian, and C.~Engblom,
  ``Managing the requirements flow from strategy to release in large-scale
  agile development: a case study at ericsson,'' \emph{Empirical Software
  Engineering}, vol.~22, no.~6, pp. 2892--2936, 2017.

\bibitem{gousios2015work}
G.~Gousios, A.~Zaidman, M.-A. Storey, and A.~Van~Deursen, ``Work practices and
  challenges in pull-based development: the integrator's perspective,'' in
  \emph{IEEE International Conference on Software Engineering}, vol.~1.\hskip
  1em plus 0.5em minus 0.4em\relax IEEE, 2015, pp. 358--368.

\bibitem{gallaba2019improving}
K.~Gallaba, ``Improving the robustness and efficiency of continuous integration
  and deployment,'' in \emph{2019 IEEE International Conference on Software
  Maintenance and Evolution (ICSME)}.\hskip 1em plus 0.5em minus 0.4em\relax
  IEEE, pp. 619--623.

\bibitem{elbaum2014techniques}
S.~Elbaum, G.~Rothermel, and J.~Penix, ``Techniques for improving regression
  testing in continuous integration development environments,'' in
  \emph{Proceedings of the 22nd ACM SIGSOFT International Symposium on
  Foundations of Software Engineering}, 2014, pp. 235--245.

\bibitem{forsgren2019accelerate}
\BIBentryALTinterwordspacing
N.~Forsgren, D.~Smith, J.~Humble, and J.~Frazelle, ``2019 accelerate state of
  devops report,'' DORA and Google Cloud, Tech. Rep. 48455, 2019. [Online].
  Available: \url{https://tinyurl.com/yauqhfqk}
\BIBentrySTDinterwordspacing

\bibitem{gallaba2018use}
K.~Gallaba and S.~McIntosh, ``Use and misuse of continuous integration
  features: An empirical study of projects that (mis) use travis ci,''
  \emph{IEEE Transactions on Software Engineering}, 2018.

\bibitem{rahman2016feature}
M.~T. Rahman, L.-P. Querel, P.~C. Rigby, and B.~Adams, ``Feature toggles:
  practitioner practices and a case study,'' in \emph{Proceedings of the 13th
  International Conference on Mining Software Repositories}, 2016, pp.
  201--211.

\bibitem{rahman2018characterizing}
A.~Rahman, A.~Agrawal, R.~Krishna, and A.~Sobran, ``Characterizing the
  influence of continuous integration,'' 2018.

\bibitem{roller2015applied}
M.~R. Roller and P.~J. Lavrakas, \emph{Applied qualitative research design: A
  total quality framework approach}.\hskip 1em plus 0.5em minus 0.4em\relax
  Guilford Publications, 2015.

\end{thebibliography}

\end{document}